%% file: paper.tex
\documentclass{llncs}

\usepackage{etex}
\usepackage{delarray}
\usepackage{amssymb}
\usepackage{amsmath}
\usepackage{graphicx}
\usepackage{subfig}
\usepackage[usenames,dvipsnames]{color}
\usepackage{url}
\usepackage[dvipsnames]{xcolor}

\usepackage{fancyvrb}

\usepackage{algorithm}
\usepackage{algorithmicx}
\usepackage{algpseudocode}

\usepackage{booktabs}
\usepackage{colortbl}

\usepackage{tikz}
\usetikzlibrary{calc,shapes,snakes,arrows,shadows}

\DeclareCaptionType{copyrightbox}

\newcommand{\INDSTATE}[1][1]{\State\hspace{#1\algorithmicindent}}

\newcommand\Ham{\hat{H}}

\newcommand\nr{{\it n}({\bf r})}
\newcommand\kv{{\bf k}}
\newcommand\rv{{\bf r}}

\newcommand\hsumpart[2]{\left(#1_{\it L'}^{a,{\bf
        G'}}\right)^{\ast}\thinspace T_{{\it L'},{\it
      L};a}^{\left[#1#2\right]}\thinspace#2_{\it L}^{a,{\bf G}}}
\usepackage{xcolor}   
\usepackage{setspace} 

\begin{document}

\title{Hybrid CPU-GPU generation of the Hamiltonian and Overlap matrices
       in FLAPW methods}

\author{Diego Fabregat-Traver$^\ast$ \and Davor Davidovi\'c$^\dagger$ \and \\ 
        Markus H\"ohnerbach$^\ast$ \and Edoardo di Napoli$^{\ddagger\ast}$}

\institute{
  $^\star$AICES, RWTH Aachen, Germany \\
  $^\dagger$RBI, Zagreb, Croatia \\
  $^\ddagger$JSC and JARA, J\"ulich, Germany \\
  \email{fabregat@aices.rwth-aachen.de,
         davor.davidovic@irb.hr, 
         hoehnerbach@aices.rwth-aachen.de, 
         e.di.napoli@fz-juelich.de}
}

\maketitle

\begin{abstract}

    In this paper we focus on the integration of high-performance numerical
    libraries in ab initio codes and the portability of performance and
    scalability. The target of our work is FLEUR, a software for electronic
    structure calculations developed in the Forschungszentrum J\"ulich over the
    course of two decades.
    The presented work follows up on a previous effort to modernize legacy code
    by re-engineering and rewriting it in terms of highly optimized libraries.
    We illustrate how this initial effort to get efficient and portable
    shared-memory code enables fast porting of the code to emerging
    heterogeneous architectures.  More specifically, we port the code to nodes
    equipped with multiple GPUs.  We divide our study in two parts. First, we
    show considerable speedups attained by minor and relatively straightforward
    code changes to off-load parts of the computation to the GPUs. Then, we
    identify further possible improvements to achieve even higher performance
    and scalability. 
    On a system consisting of 16-cores and 2 GPUs, we observe speedups of up to
    5$\times$ with respect to our optimized shared-memory code, which in turn means
    between 7.5$\times$ and 12.5$\times$ speedup with respect to the original FLEUR code.

\end{abstract}

\section{Introduction} \label{sec:intro}

\input{intro}

\section{Algorithm} \label{sec:algorithm}

As a first step towards using the BLAS and LAPACK libraries, all the involved objects
in Eqs.~\eqref{eq:def_overlap} and~\eqref{eq:def_hamilton} are expressed
in matrix form, dropping indexes $L$, $L'$, $G$, and $G'$.
Assuming the coefficient objects $A$ and $B$ as well as the $T$ matrices
as input, matrices $H$ and $S$ can be computed as follows:

\begin{align}
    H &= \sum^{N_A}_{a=1} \underbrace{A^H_a T^{[AA]} A_a}_{H_{AA}} +
                          \underbrace{
                         A^H_a T^{[AB]} B_a + 
                         B^H_a T^{[BA]} A_a + 
                         B^H_a T^{[BB]} B_a}_{H_{AB+BA+BB}} 
    \label{eq:h}
                         \\
    S &= \sum^{N_A}_{a=1} A^H_a A_a + B^H_a U^H_a U_a B_a,
    \label{eq:s}
\end{align}

\noindent
where 
$A_a$ and $B_a \in \mathbb{C}^{N_L \times N_G}$,
$T^{[...]}_a \in \mathbb{C}^{N_L \times N_L}$,
$H$ and $S \in \mathbb{C}^{N_G \times N_G}$, and
$U \in \mathbb{C}^{N_L \times N_L}$ is a diagonal matrix.
Typical for the matrix sizes are 
$N_A \sim \mathcal{O}(100)$, 
$N_G \sim \mathcal{O}(1000)$ to $\mathcal{O}(10000)$, and 
$N_L \sim \mathcal{O}(100)$.   

Algorithm~\ref{alg:hands} illustrates the algorithm used to compute
Eqs.~\eqref{eq:h} and~\eqref{eq:s} in HSDLA.  Two main insights
drive the design of the algorithm. First, it exploits symmetries to reduce the
computational cost; then, it casts the computation in terms of BLAS and LAPACK
kernels. Furthermore, when possible, multiple matrices are stacked together to allow
for larger matrix products, which in general results in higher performance.

The computation of $H$ is split into two parts, $H_{AB+BA+BB}$ and  $H_{AA}$.
The computation of $H_{AB+BA+BB}$ corresponds to lines 4 through 10. The key
idea behind the calculation is to rewrite the expression as 
$$\sum^{N_A}_{a=1} B^H_a (T^{[BA]} A_a) + (A^H_a T^{[AB]}) B_a + \frac{1}{2} B^H_a (T^{[BB]} B_a) + \frac{1}{2} (B^H_a T^{[BB]}) B_a,$$
noting that $T^{[BA]}$ is the Hermitian transpose of $T^{[AB]}$ and that $T^{[BB]}$
is itself Hermitian. The operations in parentheses are computed one at
a time for each $i$. Then, the results are aggregated into single large matrices
for a large product.

The computation of $H_{AA}$ corresponds to lines 17 through 29.  The algorithm
first attempts a Cholesky factorization of $T^{[AA]}$ ($C_a C_a = T^{[AA]}$),
which requires the matrix to be Hermitian positive definite (HPD).  While, in
theory, $T^{[AA]}$ is HPD, in practice, due to numerical considerations, the
factorization may fail. Depending on the success or failure of each individual
factorization, the results of operations in lines 21 and 24 are stacked on different
temporary operands to then compute $H_{AA}$ in two steps (lines 28 and 29).

The computation of $S$ (lines 13 through 15) is more straightforward. First,
the product $A^H A$ is computed as a single large product. Then $B$ is updated
with the norms stored in $U$ and a second large product $B^H B$ completes the
computation.

\begin{algorithm}
\small
\begin{algorithmic}[1]
    \State Create $A$, $B$
    \State Backup $\hat{A} = A$, $\hat{B} = B$ \\
    // First part of H 
    \For{$a := 1 \to N_A$}
        \State $Z_a = T^{[BA]}_a A_a$  \Comment{({\tt zgemm}: $8 N^2_L N_G$ Flops)}
        \State $Z_a = Z_a + \frac{1}{2} T^{[BB]}_a B_a$  \Comment{({\tt zhemm}: $8 N^2_L N_G$ Flops)}
        \State Stack $Z_a$ to $Z$
        \State Stack $B_a$ to $B$
    \EndFor
    \State $H = Z^H B + B^H Z$         \Comment{({\tt zher2k}: $8 N_A N_L N^2_G$ Flops)}
    \State Restore $A = \hat{A}$, $B = \hat{B}$ \\
    // S 
    \State $S = A^H A$      \Comment{({\tt zherk}: $4 N_A N_L N^2_G$ Flops)}
    \State $B = U B$        \Comment{({\tt scaling}: $2 N_A N_L N_G$ Flops)}
    \State $S = S + B^H B$  \Comment{({\tt zherk}: $4 N_A N_L N^2_G$ Flops)}\\
    // Second part of H
    \For{$a := 1 \to N_A$}
        \State {\bf try:}
        \INDSTATE[0] $C_a = Cholesky(T^{[AA]}_a)$    \Comment{({\tt zpotrf}: $\frac{4}{3} N^3_L$ Flops)}
        \State {\bf success:}
        \INDSTATE[0] $Y_a = C^H_a A_a$               \Comment{({\tt ztrmm}: $4 N^2_L N_G$ Flops)}
        \INDSTATE[0] Stack $Y_a$ to $Y_{\text{HPD}}$
        \State {\bf failure:}
        \INDSTATE[0] $X_a = T^{[AA]}_a A_a$          \Comment{({\tt zhemm}: $8 N^2_L N_G$ Flops)}
        \INDSTATE[0] Stack $X_a$ to $X_{\neg \text{HPD}}$
        \INDSTATE[0] Stack $A_a$ to $A_{\neg \text{HPD}}$
    \EndFor
    \State $H = H + A^H_{\neg \text{HPD}} X_{\neg \text{HPD}}$  \Comment{({\tt zgemm}: $8 N_{A_{\neg \text{HPD}}} N_L N^2_G$ Flops)}
    \State $H = H + Y^H_{\text{HPD}} Y_{\text{HPD}}$            \Comment{({\tt zherk}: $4 N_{A_{\text{HPD}}} N_L N^2_G$ Flops)}
    
\end{algorithmic}
    \caption{{\bf: Computation of the H and S matrices in HSDLA.}}
    \label{alg:hands}
\end{algorithm}

\section{Software re-engineering and performance portability}
\label{sec:portability}

In this section we set the focus on the porting of the multi-core
implementation of Alg.~\ref{alg:hands} to heterogeneous architectures
consisting of one multi-core node equipped with one or more GPUs. 
We perform a quick analysis of the computation to determine how to
split the computation between CPU and GPU(s) with minimal modifications
to the code, and illustrate how with these minor modifications one can 
already benefit considerably from the combined computational power of CPU and GPUs.

Given the characteristic values for $N_A$, $N_L$, and $N_G$ observed
in our test cases, at least 97\% of the computation is performed
by the operations in lines 10, 13, 15, 28 and 29. 
Thanks to the aggregation of many small matrix products into single
large ones, all these 5 operations are large enough to be efficiently
computed on the GPUs.  Therefore, the first step is to off-load these
computations to the GPU making sure that relatively high efficiency is
attained.

All five calls correspond to BLAS kernels; we thus look into available
BLAS implementations for GPUs.  There exists a range of GPU libraries
that offer BLAS functionality, both academic and commercial, such as
cuBLAS~\cite{CUBLAS}, cuBLAS-XT, MAGMA~\cite{tnld10}, and
BLASX~\cite{Wang2015:BLASX}. The first two are commercial and
developed by NVIDIA, the other two are academic efforts. From the
point of view of programmability, the most convenient alternatives are
cuBLAS-XT and BLASX, since they require minor or no changes to the
calls to BLAS routines and take also care of the data transfers
between CPU and GPU transparently. While BLASX offers certain
advantages from the programmability perspective and claims higher
performance and scalability (see~\cite{Wang2015:BLASX}), we
encountered some problems in the integration and opted for using
cuBLAS-XT for our initial study.

Since cuBLAS-XT does not abide to the BLAS standard interface, three wrappers, of
about 15 lines of code each, around the calls to {\tt zherk}, {\tt zher2k} and {\tt
zgemm} are required to ensure the code works seamlessly in both CPU-only and CPU-GPU(s)
modes. An example for {\tt zgemm} follows: 

\begin{Verbatim}[frame=lines,fontsize=\scriptsize]
void gpu_zgemm_( char *transa, char *transb, int *m,  int *n, int *k,
                 std::complex<double> *alpha,  std::complex<double> *A,  int *lda,
                 std::complex<double> *B,  int *ldb, 
                 std::complex<double> *beta, std::complex<double> *C,  int *ldc )
{
    cublasOperation_t cu_transa = transa[0] == 'N' ? CUBLAS_OP_N : 
                                  transa[0] == 'T' ? CUBLAS_OP_T : CUBLAS_OP_C;
    cublasOperation_t cu_transb = transb[0] == 'N' ? CUBLAS_OP_N : 
                                  transb[0] == 'T' ? CUBLAS_OP_T : CUBLAS_OP_C;
    cublasXtZgemm( handle, cu_transa, cu_transb, *m, *n, *k, 
                   (cuDoubleComplex *)alpha, (cuDoubleComplex *)A, *lda,
                                             (cuDoubleComplex *)B, *ldb,
                   (cuDoubleComplex *)beta,  (cuDoubleComplex *)C, *ldc );
}
\end{Verbatim}

\noindent
In addition, two routines for proper initialization and cleanup of the
cuda runtime and the devices are needed. Finally, for the data transfers
between CPU and GPU to be efficient, memory for the matrices involved
must be pinned (page-locked).

With these minor modifications, about 100 lines of extra coding,
the program is ready to off-load most of the computation to 
multiple GPUs and attain n-fold speedups.
It is important to highlight that this simple extension is only
possible thanks to the initial effort in rewriting the initial
FLEUR code in terms of matrix (BLAS/LAPACK) operations. At
that point the efficiency and scalability of the code may
be easily ported to more complex architectures.
Had the original FLEUR code not undergone the reengineering
process, the coding of efficient low level routines for
the GPUs would be a much more complex and time-consuming effort.

\section{Experimental Results} \label{sec:experiments}

We turn now our attention to experimental results. We compare the performance
of our hybrid CPU-GPU implementation of Alg.~\ref{alg:hands} with
the performance of the multi-core (CPU only) HSDLA.
As test cases we use two input systems
describing two distinct physical systems, to which we refer as NaCl
and AuAg, respectively. By including both an insulator and a
conductor, these systems represent a heterogeneous sample with
different physical properties. The tests generate the matrices $H$ and
$S$ for one single \kv -point, and different $K_{max}$ values. The actual
problem sizes, that is, the values for $N_A$, $N_L$, and $N_G$ in each
case are given in Tab.~\ref{tab:sizes}.

\begin{table}
    \centering
    \begin{tabular}{ @{\hspace{1mm}} c @{\hspace{1mm}} | @{\hspace{2mm}} c 
                     @{\hspace{3mm}} c @{\hspace{2mm}} | @{\hspace{2mm}} c 
                     @{\hspace{3mm}} c @{\hspace{3mm}} c 
                     @{\hspace{3mm}} c @{\hspace{3mm}} c @{\hspace{1mm}}} \toprule
        {\bf Test case} & $N_A$ & $N_L$ & $N_G:$ & 
                          {\scriptsize $K_{max}=2.5$} & 
                          {\scriptsize $K_{max}=3.0$} & 
                          {\scriptsize $K_{max}=3.5$} & 
                          {\scriptsize $K_{max}=4.0$} \\ \midrule
        NaCl      & 512 &  49 & & 2256 &  3893 &  6217 &  9273 \\
        AuAg      & 108 & 121 & & 3275 &  5638 &  8970 & 13379 \\\bottomrule
    \end{tabular}
    \caption{Problem sizes for NaCl and AuAg and for a variety of $K_{max}$ values.
             The value of $N_G$ varies with $K_{max}$.}
    \label{tab:sizes}
\end{table}

We ran our experiments in two different compute nodes, which we will
refer to as RWTH and JURECA. 
The RWTH node consists of two eight-core Sandy Bridge E5-2650 processors, running
at a nominal frequency of 2.0 GHz, and 2 NVIDIA Tesla K20Xm GPUs. The node is
equipped with 64 GBs of RAM. 
The combined peak performance for the 16 CPU cores in double precision is of
256 GFlops, while the peak performance for double precision of each GPU is of
1.3 TFlops, for a total of 2.6 TFlops.
The JURECA node consists of two twelve-core Haswell E5-2680v3 processors, running
at a nominal frequency of 2.5 GHz, and 2 NVIDIA K80 GPUs (each of which
consists of two K40 GPU devices). The node is equipped with 128 GBs of
RAM.
The combined peak performance for the 24 CPU cores in double precision is of
960 GFlops, while the peak performance for double precision of each GPU device
is of about 1.45 TFlops, for a total of 5.8 TFlops.
In all cases, the code was linked to Intel MKL version 11.3.2 for the BLAS
and LAPACK routines on the CPU; the GPU code was linked to NVIDIA cuBLAS-XT version 7.5.

\paragraph{\bf RWTH.}

Table~\ref{tab:nacl} collects the timings for the NaCl test case for the three
scenarios of interest (CPU only, CPU + 1 GPU, CPU + 2 GPUs). The speedup with respect
to HSDLA is given in parentheses. As expected, the considerable gap in performance
between CPU and GPU is reflected in the observed large speedups:
up to $2.76\times$ and $4.04\times$ for 1 and 2 GPUs, respectively. 

Similar results are presented in Tab.~\ref{tab:auag} for the AuAg
test case, but in this case the observed speedups are even larger. 
The reason for this is that, while MKL is already close to its peak performance
for the matrix sizes of NaCl, cuBLAS-XT still has room for improvement and
benefits from the larger matrices in AuAg. In fact, 
one can expect still better speedups for larger systems.

\begin{table}
    \centering
    \begin{tabular}{ @{\hspace{1mm}} l @{\hspace{1mm}} |
                     @{\hspace{3mm}} r @{\hspace{4mm}} r
                     @{\hspace{4mm}} r @{\hspace{4mm}} r @{\hspace{1mm}}} \toprule
        {\bf Setup} & {\footnotesize $K_{max}=2.5$} & 
                      {\footnotesize $K_{max}=3.0$} & 
                      {\footnotesize $K_{max}=3.5$} & 
                      {\footnotesize $K_{max}=4.0$} \\ \midrule
        CPU only      & 18.27s \phantom{(2.28$\times$)}  & 39.84s \phantom{(2.28$\times$)}    & 91.52s \phantom{(2.28$\times$)}    & 189.53s \phantom{(2.28$\times$}    \\
        CPU + 1 GPU   &  8.03s (2.28$\times$) & 15.87s (2.51$\times$) & 35.64s (2.57$\times$) &  68.59s (2.76$\times$) \\
        CPU + 2 GPUs  &  6.51s (2.81$\times$) & 12.37s (3.22$\times$) & 24.39s (3.75$\times$) &  46.97s (4.04$\times$) \\\bottomrule
    \end{tabular}
    \caption{Timings and speedup (in parentheses) for the NaCl test case for varying $K_{max}$.
    Results for the RWTH node.}
    \label{tab:nacl}
\end{table}
\vspace{-10mm}

\begin{table}
    \centering
    \begin{tabular}{ @{\hspace{1mm}} l @{\hspace{1mm}} |
                     @{\hspace{3mm}} r @{\hspace{4mm}} r
                     @{\hspace{4mm}} r @{\hspace{4mm}} r @{\hspace{1mm}}} \toprule
        {\bf Setup} & {\footnotesize $K_{max}=2.5$} & 
                      {\footnotesize $K_{max}=3.0$} & 
                      {\footnotesize $K_{max}=3.5$} & 
                      {\footnotesize $K_{max}=4.0$} \\ \midrule
        CPU only      & 15.64s \phantom{(2.28$\times$)} & 46.23s \phantom{(2.28$\times$)} & 104.25s \phantom{(2.28$\times$)} & 215.98s \phantom{(2.28$\times$)}    \\
        CPU + 1 GPU   &  7.52s (2.08$\times$) & 16.16s (2.86$\times$) & 35.62s (2.93$\times$) & 71.35s (3.03$\times$) \\
        CPU + 2 GPUs  &  5.62s (2.78$\times$) & 11.28s (4.10$\times$) & 23.10s (4.51$\times$) & 43.54s (4.96$\times$) \\\bottomrule
    \end{tabular}
    \caption{Timings and speedup (in parentheses) for the AuAg test case for varying $K_{max}$.
    Results for the RWTH node.}
    \label{tab:auag}
\end{table}
\vspace{-10mm}

\paragraph{\bf JURECA.}
Results for the JURECA node are presented in
Tabs.~\ref{tab:nacl-jureca} and~\ref{tab:auag-jureca} for NaCl and
AuAg, respectively. In this case we show timings and speedups for up
to 4 GPUs. The maximum observed speedups are $1.77\times$,
$2.76\times$ and $4.26\times$ for 1, 2 and 4 GPUs, respectively. Given
that the increase in computational power in each case is of
$2.4\times$, $3.9\times$ and $6.8\times$, respectively, these numbers
are quite satisfactory.

\begin{table}
    \centering
    \begin{tabular}{ @{\hspace{1mm}} l @{\hspace{1mm}} |
                     @{\hspace{3mm}} r @{\hspace{4mm}} r
                     @{\hspace{4mm}} r @{\hspace{4mm}} r @{\hspace{1mm}}} \toprule
        {\bf Setup} & {\footnotesize $K_{max}=2.5$} & 
                      {\footnotesize $K_{max}=3.0$} & 
                      {\footnotesize $K_{max}=3.5$} & 
                      {\footnotesize $K_{max}=4.0$} \\ \midrule
        CPU only      & 9.334s \phantom{(2.28$\times$)} & 23.293s \phantom{(2.28$\times$)} & 41.500s \phantom{(2.28$\times$)} & 74.731s \phantom{(2.28$\times$)}    \\
        CPU + 1 GPU   & 6.474s (1.44$\times$) & 14.502s (1.61$\times$) & 32.728s (1.27$\times$) & 66.546s (1.12$\times$) \\
        CPU + 2 GPUs  & 4.995s (1.87$\times$) & 10.381s (2.24$\times$) & 21.842s (1.90$\times$) & 42.581s (1.76$\times$) \\
        CPU + 4 GPUs  & 4.720s (1.98$\times$) &  8.760s (2.66$\times$) & 15.449s (2.69$\times$) & 26.575s (2.81$\times$) \\\bottomrule
    \end{tabular}
    \caption{Timings and speedup (in parentheses) for the NaCl test case for varying $K_{max}$.
    Results for the JURECA node.}
    \label{tab:nacl-jureca}
\end{table}

\begin{table}
    \centering
    \begin{tabular}{ @{\hspace{1mm}} l @{\hspace{1mm}} |
                     @{\hspace{3mm}} r @{\hspace{4mm}} r
                     @{\hspace{4mm}} r @{\hspace{4mm}} r @{\hspace{1mm}}} \toprule
        {\bf Setup} & {\footnotesize $K_{max}=2.5$} & 
                      {\footnotesize $K_{max}=3.0$} & 
                      {\footnotesize $K_{max}=3.5$} & 
                      {\footnotesize $K_{max}=4.0$} \\ \midrule
        CPU only      & 9.102s \phantom{(2.28$\times$)} & 22.681s \phantom{(2.28$\times$)} & 57.314s \phantom{(2.28$\times$)} & 100.190s \phantom{(2.28$\times$)}    \\
        CPU + 1 GPU   & 6.136s (1.48$\times$) & 14.466s (1.57$\times$) & 32.338s (1.77$\times$) & 68.910s (1.45$\times$) \\
        CPU + 2 GPUs  & 4.376s (2.08$\times$) &  9.699s (2.34$\times$) & 20.800s (2.76$\times$) & 42.242s (2.37$\times$) \\
        CPU + 4 GPUs  & 3.533s (2.58$\times$) &  6.690s (3.39$\times$) & 13.457s (4.26$\times$) & 25.359s (3.95$\times$) \\\bottomrule
    \end{tabular}
    \caption{Timings and speedup (in parentheses) for the AuAg test case for varying $K_{max}$.
    Results for the JURECA node.}
    \label{tab:auag-jureca}
\end{table}

\subsection{Fine-tuning for performance and scalability}
\label{sec:tuning}

The observed speedups are substantial.
Yet, one could expect even better results, especially in the case of
the RWTH node where the computational power of the two GPUs combined is ten
times larger than that of the CPUs.  This potential for improvement
comes as no surprise, since this is only a basic port to illustrate how far
one can get with minimal code modifications; to attain close-to-optimal
performance further work is required.  For ideal results, a hybrid and
highly-tuned BLAS as well as a GPU-accelerated version of the computation in
the loops are needed.

In order to have a more tangible discussion, we provide in
Tab.~\ref{tab:sample-results} a breakdown of the timings for the NaCl
($K_{max} = 4.0$) test case running in the RWTH node with two K20x GPUs. The bottom rows correspond to the large BLAS
operations off-loaded to the two GPUs; the top rows correspond to the rest of
the code (both loops and the application of the $U$ norm) executed on the CPU
only. 
The efficiency is measured with respect to the combined performance of CPU plus GPUs.

\begin{table}
    \centering
    \begin{tabular}{ @{\hspace{3mm}}l @{\hspace{3mm}} | @{\hspace{3mm}} c @{\hspace{7mm}} r @{\hspace{5mm}} c @{\hspace{3mm}}} \toprule
        {\bf Section (line(s))} & {\bf Time} & {\bf Performance} & {\bf Efficiency} \\ \midrule
        Loop 1 (4--9)   &   2.27 secs  &    80.35 GFlops/s  &  0.03 \\
        Loop 2 (17--27) &   2.62 secs  &    34.81 GFlops/s  &  0.01 \\
        U norm (14)    &   0.23 secs  &     1.01 GFlops/s  &  0.00 \\ \midrule
        S1 (13) &   4.37 secs  &  1974.63 GFlops/s  &  0.69 \\
        S2 (15) &   4.41 secs  &  1956.72 GFlops/s  &  0.68 \\ 
        H1 (10) &   9.49 secs  &  1818.57 GFlops/s  &  0.63 \\
        H2 (28) &   2.32 secs  &  1859.72 GFlops/s  &  0.65 \\
        H3 (29) &   4.75 secs  &  1816.66 GFlops/s  &  0.63 \\ \bottomrule
    \end{tabular}
    \caption{Breakdown of timings for NaCl ($K_{max} = 4.0$) together with the respective
             attained performance and efficiency.} 
    \label{tab:sample-results}
\end{table}

Three main messages can be extracted from Tab.~\ref{tab:sample-results}:
\begin{enumerate}
    \item NVIDIA's cuBLAS-XT does a good job 
        attaining  an efficiency between $63\%$ and $69\%$.
    \item Yet, these operations may attain about $90\%$ of the peak if the
        matrices are large enough and the code is highly optimized and hybrid.
        This would mean attaining around 2.5 TFlops/s, that is, an extra
        $25\%$ speedup for these heavy computations.
    \item When the target architecture offers massive parallelism, minor portions of code that
        do not scale may become a bottleneck. In our case, the $3\%$ of the 
        computation that was not off-loaded to the GPUs becomes non-negligible.
        In fact, the weight of these operations in our experiments may account for
        up to 35\% of the time to solution, and compromise the overall scalability.
        Due to the size of the matrices involved in these operations
        (between $50 \times 50$ and $100 \times 100$ for the $T$ matrices
        in our test cases), these products do not scale well, especially on GPUs.
        Specialized code is required to mitigate their impact in the overall time
        to solution.
\end{enumerate}

\section{Conclusions and Future Work} \label{sec:conclusions}

We concentrated on the benefits of rewriting scientific code in terms of
standardized libraries for portable performance and scalability. As use
case we considered a portion of the FLEUR code base, a software for electronic
structure calculations.

We demonstrated that major efforts in re-engineering part of the
original FLEUR code, and writing it in terms of the BLAS and LAPACK
libraries, enables a fast porting that exploits the vast computational
power of emerging heterogeneous architectures such as multi-core CPUs
combined with multiple GPUs. Most importantly, the porting only
required less than 100 new lines of code.  The resulting
implementation attains speedups of up to $3\times$ and $5\times$ for
simulations run on a system equipped with two K20x GPUs, respectively,
and speedups of up to $1.8\times$, $2.8\times$ and $4.3\times$ for
runs with 1, 2 and 4 GPUs, respectively, on a system equipped with two
K80 GPUs (each consisting of two K40 GPUs).

While satisfactory, these results highlight room for improvement. In the future,
we aim at developing more efficient hybrid CPU-GPU routines for the major
matrix products in the code as well as attaining sufficient scalability of the
rest of the code to ensure a uniform overall scalability.

\section{Acknowledgements}

This work was partially funded by 
the Ministry of Science and Education of the Republic of Croatia and
the Deutsche Akademische Austauschdienst (DAAD) from funds of the 
Bundesministeriums f\"ur Bildung und Forschung (BMBF) through project 
``PPP Kroatien'' ID 57216700.
Financial support from the J\"ulich Aachen Research Alliance-High Performance
Computing and the Deutsche Forschungsgemeinschaft (DFG) through grant GSC 111 is
also gratefully acknowledged. Furthermore, the authors thank the RWTH IT Center
and the J\"ulich Supercomputing Centre for the computational
resources.


\end{document}

%% file: intro.tex



Many legacy codes in scientific computing have grown over time with an eye on
functionality, but little emphasis on portable performance and scalability. Often, 
these codes are a direct translation of mathematical formulae, and lack
proper engineering (i.e. modularity, code reuse, etc). One such example is FLEUR, 
a software for electronic structure calculations developed at the J\"ulich
Research Center during the last two decades~\cite{FLEUR}.
In previous work by Di Napoli et al.~\cite{DiNapoli:HSDLA}, the
authors made the effort of reengineering a portion of FLEUR's code base
in an attempt to demonstrate the value of writing the
computational bottlenecks in terms of kernels provided by standardized
and highly-tuned libraries. There, they show an increase in
performance and anticipate its portability beyond multi-core
architectures. In this paper, we confirm that the reengineering
process indeed guarantees quick portability and high-performance on
emerging heterogeneous architectures, as in the case of multi-core
CPUs equipped with one or more coprocessors such as GPUs.

In times where massively-parallel heterogeneous architectures have
become the most common computing platform, legacy scientific code has
to be modernized.  New software is often designed with portable
efficiency and scalability in mind, and some old code is undergoing
major rewriting to adapt to the newest architectures. However, there
is still a lot of reluctance to undergo through the rewriting process
since it requires a vast initial effort and may incur in validation
issues. While it is understandable that domain scientists are hesitant
to introduce major changes into a code developed and tested in the
course of many years, legacy codes that do not go through this process
are destined to be marginalized.


%
%

A critical insight in writing long-lasting scientific code is to have
a modular design where, at the bottom layers, the computational
bottlenecks are written in terms of kernels available from
standardized and highly-tuned libraries. Examples of such kernels are
fast Fourier transforms, matrix products, and eigensolvers, provided
by a number of commercial as well as academic libraries.
The most prominent example of standard and optimized scientific
library is the Basic Linear Algebra Subprograms (BLAS). This library
has its roots in the early realization of the necessity for portable
performance. Today, BLAS kernels, which include matrix products and
linear systems, are the building blocks for a multitude of libraries,
so much that BLAS is the first library to be ported to and optimized
for every new architecture. Therefore, writing code on top of BLAS and
other standardized and broadly available libraries constitutes a first
and essential step in the modernization of scientific software.

%

In this paper we follow the same approach illustrated
in~\cite{DiNapoli:HSDLA}, where the authors made a major effort to
address the computational bottlenecks of the FLEUR's code base: the
generation of the so-called Hamiltonian and Overlap matrices. In
generating such matrices, the main goal of the original FLEUR code was the
minimization of memory usage. Furthermore there is no notion of
abstraction and encapsulation, and the different modules are tightly
coupled. At this point, low-level optimizations were unfeasible, and
the authors opted for a clean slate approach: starting from the
mathematical formulation behind the code, they cast it in terms of
matrix operations supported by the BLAS and LAPACK libraries. As
presented in their results, despite lacking some mathematical insight
that reduced the amount of computation in the FLEUR code, HSDLA (the new code)
outperformed the original one with speedups between $1.5\times$ and
$2.5\times$. Most importantly, the authors claim that  HSDLA  could be easily
ported to other architectures.

In this paper, we continue their work, and illustrate how such an
initial reengineering effort enables a quick port to heterogeneous
architectures consisting of multi-core CPUs and one or more GPUs.
More specifically, we quantify the minimal effort required in terms of
additional code to attain substantial speedups. When running on a
system equipped with two GPUs, we observe speedups of up to $5\times$
with respect to HSDLA
and about one order of magnitude with respect to the corresponding
FLEUR code.  Moreover, we identify the additional work
required to attain close-to-optimal efficiency and scalability, and
partially implement it to illustrate the idea. Finally, beyond the
results specific to the case of FLEUR, the main contribution of this
paper is to demonstrate that, despite the major initial effort, a
reengineering of legacy codes is not only worth it but imperative in
order to obtain long-lasting portable performance and scalability.

This paper is organized as follows. Section~\ref{sec:flapw} provides
the background on Density Functional Theory (DFT) and the math behind
the computation to generate the Hamiltonian and Overlap matrices.
Section~\ref{sec:algorithm} gives an overview of the optimized
algorithm behind HSDLA for the generation of
these matrices.  In Sec.~\ref{sec:portability}
we discuss the porting of the code to heterogeneous architectures,
including a review of the available BLAS libraries for GPUs and a
simple analysis of the computation to decide which portions of the
code to off-load to the GPUs.  Section~\ref{sec:experiments} presents
experimental results for 1, 2 and 4 GPUs, and points at potential improvements to the hybrid code. Finally, Sec.~\ref{sec:conclusions} draws conclusion and discusses future
research directions.



\subsection{DFT, FLAPW and the H and S matrices}
\label{sec:flapw}

The FLEUR code is based on the widely accepted framework of
Density Functional Theory (DFT).
In the last decade, Density Functional Theory (DFT)~\cite{Nogueira:1391332,Sholl:2011td} has
become the ``standard model'' in Materials Science. Within the DFT
framework, it is possible to simulate the physical properties of
complex quantum mechanical systems made of few dozens up to few
hundreds of atoms. The core of the method relies on the simultaneous
solution of a set of Schr\"odinger-like equations. These equations are
determined by a Hamiltonian operator $\Ham$ containing an effective
potential $V_0[n]$ that depends functionally on the one-particle
electron density $\nr$. In turn, the solutions of the equations
$\psi_i({\bf r})$ determine the one-particle electron
density $\nr$ used in calculating the effective potential $V_0$. 
\begin{eqnarray}
\label{eq:KSeq}
\begin{array}[l]{l}
		\Ham \psi_i({\bf r}) = \left( -\frac{\hbar^2}{2m} \nabla^2 + V_0({\bf r}) \right) \psi_i({\bf r}) = \epsilon_i \psi_i({\bf r}) \quad ; \quad \epsilon_1 \leq \epsilon_2 \leq \dots\\[2mm]
		\nr = \sum_i^N |\psi_i({\bf r})|^2
\end{array}
\end{eqnarray}

In practice, this set of equations, also known as Kohn-Sham
(KS)~\cite{Kohn:1965zzb}, is solved self-consistently; an initial guess for
$n_0({\bf r})$ is used to calculate the effective potential $V_0$ which, in
turn, is inserted in Eq.~\eqref{eq:KSeq} whose solutions,
$\psi_i({\bf r})$, are used to compute a new charge density
$n_1({\bf r})$. Convergence is checked by comparing the new density to the
starting one. When convergence is not reached, an opportune mixing of
the two densities is selected as a new guess, and the process is
repeated. This is properly called a Self-Consistent Field (SCF)
iteration.

\sloppypar
In principle, the theory only requires as input the quantum numbers
and the positions of the atoms that are part of the investigated
system. In practice, there is a plethora of DFT methods which depends
on the {\em discretization} used to parameterize the KS equations. The
discretization in the Full-potential Linearized Augmented Plane Wave
(FLAPW) method~\cite{Wimmer:1981hd,Jansen:1984bc} is based on plane wave expansion of
$\psi_{\kv,\nu}(\rv)$, where the momentum vector $\kv$ and the band
index $\nu$ replace the generic index $i$. The \kv-point wave function
$\psi_{\kv,\nu}(\rv) = \sum_{|{\bf G + k}|\leq {\bf K}_{max}} c^{\bf
  G}_{\kv,\nu} \varphi_{\bf G}(\kv,\rv)$
is expanded in terms of a basis set $\varphi_{\bf G}(\kv,\rv)$ indexed
by the vectors ${\bf G}$ lying in the lattice reciprocal to
configuration space up to a chosen cut-off value ${\bf K}_{max}$. In
FLAPW, the physical (configuration) space of the simulation cell is
divided into spherical regions, called Muffin-Tin (MT) spheres,
centered around atomic nuclei, and interstitial areas between the MT
spheres. The basis set $\varphi_{\bf G}(\kv,\rv)$ takes a different
expression depending on the region
\begin{eqnarray}
\label{eq:basis}
\varphi_{\bf G}(\kv,\rv) \propto \left\{
	\begin{array}[l]{lr}
	e^{i({\bf k+G})\rv} & \quad \textrm{\small Interstitial}\\
	\displaystyle\sum_{\it l,m} \left[A^{a,{\bf G}}_{\it l,m}(\kv) u^{a}_{\it l}(r) 
	+ B^{a,{\bf G}}_{\it l,m}(\kv) \dot{u}^{a}_{\it l}(r)
          \right] Y_{\it l,m}(\hat{\bf r}_{a}) & \quad
        \textrm{\small $a^{th}$ Muffin Tin}\\
	\end{array}
\right.
\end{eqnarray}
where the coefficients $A^{a,{\bf G}}_{\it l,m}(\kv)$ and
$B^{a,{\bf G}}_{\it l,m}(\kv)$ are determined by imposing
continuity of $\varphi_{\bf G}(\kv,\rv)$ and its derivative at the
boundary of the MTs. Due to this expansion the KS equations naturally
translate to a set of generalized eigenvalue problems
$\sum_{\bf G'} \left[ H_{\bf G,G'}(\kv) - \lambda_{\kv\nu} S_{\bf
    G,G'}(\kv) \right] c^{\bf G'}_{\kv ,\nu} =0$
for the coefficients of the expansion $c^{\bf G'}_{\kv ,\nu}$ where the
Hamiltonian and Overlap matrices $H$ and $S$ are given by multiple
integrals and sums 
\begin{equation}
\label{eq:HSdef} 
\{H(\kv),S(\kv)\}_{\bf G,G'} = \sum_a \iint \varphi^{\ast}_{\bf
  G}(\kv,\rv) \{\Ham,I\} \varphi_{\bf G'}(\kv,\rv) {\rm d} \rv.
\end{equation}
Since the set of basis functions in Eq.~\eqref{eq:basis} is
implicitly labeled by the values the variable $\kv$ takes in the
Brillouin zone, there are multiple Hamiltonian and Overlap matrices,
one for each independent $\kv$-point.

Without loss of generality, we can abstract from the \kv -point index
and recover an explicit formulation of the $H_{\bf G,G'}$ and
$S_{\bf G,G'}$ matrices by substituting Eq.~\eqref{eq:basis} in
Eq.~\eqref{eq:HSdef} and carrying out the multiple integration. The
computation is particularly complex within the MT regions where the
initialization of the Hamiltonian and Overlap matrices is by far the
most computationally intensive task. By exploiting the properties of
the basis functions, the $H$ and $S$ matrices are directly expressed
as functions of the set of $A$ and $B$ coefficients.
\begin{equation}
\left(S\right)_{\bf G',G}
	=\sum_{a=1}^{N_A}\sum_{{\it l,m}}
     \left(A^{a,{\bf G'}}_{\it l,m}\right)^{\ast}A_{\it l,m}^{a,{\bf G}}
    +\left(B_{\it l,m}^{a,{\bf G'}}\right)^{\ast}B_{\it l,m}^{a,{\bf G}} \left\|
      \dot u_{\it l}^a\right\|^2
\label{eq:def_overlap}
\end{equation}
\begin{align}
\left(H\right)_{G',G}=\sum_{a=1}^{N_A}\sum_{L',L} & \left(\hsumpart AA\right)+\left(\hsumpart AB\right)\nonumber \\
+ & \left(\hsumpart BA\right)+\left(\hsumpart BB\right).
\label{eq:def_hamilton}
\end{align}
Notice that in Eq.~\eqref{eq:def_hamilton} for convenience we have
compacted the indexes ${\it l,m}$ into ${\it L}$, and expressed the
range of the index $a$ over all the distinct atom types $N_A$.  The
new matrices
$T_{L',L;a}^{\left[\dots\right]}\in \mathbb C^{N_L\times N_L}$ are
dense and their computation involves multiple integrals
between the basis functions and the non-spherical part of the
potential $V_0$ (See \cite[Appendix A.2]{DiNapoli:HSDLA} for details).
Due to the non-orthornormality of the basis function set
\eqref{eq:basis}, the matrix $S$ is non-diagonal, dense, and
generically positive definite with the exception of having few very
small singular values. On the opposite $H$ is always non-definite and
both matrices are either complex Hermitian or real symmetric.